\documentclass[aps,prl,preprint,groupedaddress]{revtex4}

\usepackage{graphicx}
\usepackage{dcolumn}
\usepackage{bm}

\begin{document}

\preprint{LA-UR-05-5164}

\title{Allostery in a Coarse-Grained Model of Protein Dynamics}



\author{Dengming Ming}
\affiliation{Computer and Computational Sciences Division, Los Alamos National Laboratory, Los Alamos, NM 87545, USA}

\author{Michael E. Wall}
\email[Correspondence: ]{mewall@lanl.gov}
\affiliation{Computer and Computational Sciences and Bioscience Divisions, Los Alamos National Laboratory, Los Alamos, NM 87545, USA}


\date{\today}

\begin{abstract}

We propose a criterion for optimal parameter selection in
coarse-grained models of proteins, and develop a refined elastic
network model (ENM) of bovine trypsinogen. The unimodal density-of-states
distribution of the trypsinogen ENM disagrees with the bimodal
distribution obtained from an all-atom model; however, the bimodal
distribution is recovered by strengthening interactions between atoms
that are backbone neighbors. We use the backbone-enhanced model to
analyze allosteric mechanisms of trypsinogen, and find relatively strong
communication between the regulatory and active sites.
\end{abstract}


\maketitle


A major challenge of molecular biology is to understand regulatory
mechanisms in large protein complexes that are abundant in
multi-celluluar organisms. To make simulation of such complexes
computationally feasible, coarse-grained models have been developed,
in which a subset of the atoms in the complex are used to simulate the
large-scale motions. However, principled methods to quantify and
optimize the accuracy of coarse-grained models are currently lacking.

In one common coarse-graining method, an all-atom model is simplified
by considering effective interactions among a subset of the atoms
(e.g., just the alpha-carbons). The usual criterion for model accuracy
is the ability of a model to reproduce atomic mean-squared
displacements (MSDs). However, MSDs are just one aspect of protein
dynamics -- a stricter criterion for the accuracy of a coarse-grained
model is the similarity between the configurational distributions of
the selected atoms in the coarse-grained and all-atom models. Such a
criterion is also biologically relevant, in part because the
conformational distribution is a key determinant of protein activity
\cite{Frauenfelder85}.

One useful measure of the difference between conformational
distributions is the Kullback-Leibler divergence $D_{\bf x}$ (see
definition below) \cite{Kullback51,Ming05}. Recently, an analytic
expression for $D_{\bf x}$ was obtained for harmonic vibrations of a
protein-ligand complex both with and without a protein-ligand
interaction \cite{Ming05}. Here we show how an equivalent expression
may be applied to refine a coarse-grained model of protein
dynamics. To use the expression for $D_{\bf x}$ requires the marginal
probability distribution of a subset of the atoms in a protein, which
we calculate in the harmonic approximation. We then apply the
equations to refine an anisotropic elastic network model (ENM)
\cite{Atilgan01} of trypsinogen dynamics with respect to an all-atom
model calculated using CHARMM \cite{Brooks83}. The unimodal
density-of-states distribution of the ENM disagrees with the bimodal
distribution obtained from the all-atom model; however, the bimodal
distribution is recovered by strengthening interactions between atoms
that are backbone neighbors. Finally, the backbone-enhanced elastic
network model (BENM) is used to analyze allosteric mechanisms of
trypsinogen, revealing relatively strong communication between the
regulatory and active sites.

Let $P({\bf x})$ be the probability distribution of the $3N$ atomic
coordinates ${\bf x}=(x_1,y_1,z_1,\ldots,x_N,y_N,z_N)$ of a molecular
model in the harmonic approximation. Let ${\bf x}=({\bf x}_1,{\bf
x}_2)$, where ${\bf x}_1$ is the $3N_1$ coordinates of a subset of
atoms of interest, and ${\bf x}_2$ is the $3N_2$ coordinates of the
remaining atoms. We are interested in calculating the marginal
distribution $P({\bf x}_1)$:
\begin{equation}
P({\bf x}_1) = \int d^{3N_2} {\bf x}_1 \, P({\bf x}_1, {\bf x}_2).
\label{eq:px1_general}
\end{equation}

We now calculate $P({\bf x}_1)$ in a model of molecular vibrations. Consider a harmonic approximation to the potential energy function
$U({\bf x})$, where ${\bf x}$ is the deviation from an
equilibrium conformation ${\bf x}_0$:
\begin{equation}
U({\bf x} + {\bf x}_0)\approx U({\bf x}_0)+{1 \over 2}{\bf x}^\dag {\bf H}\,{\bf x}.
\end{equation}
The matrix ${\bf H}$ is the Hessian of $U$ evaluated at ${\bf x}_0$: $H_{ij}|_{{\bf x}_0}=\partial^2 U / \partial x_i \partial x_j |_{{\bf x}_0}.$ We assume a Boltzmann distribution for $P({\bf x})$, and ignore solvent and pressure effects:
\begin{equation}
P({\bf x})=Z^{-1}e^{{-{\bf x}^\dag {\bf H}\,{\bf x}}\over{2 k_B T}}=(2\pi k_B T)^{-3N/2}e^{{-\left|{\bf \Omega} {\bf V}^\dag {\bf x}\right|^2} \over {2 k_B T}}\prod_{i=1}^{3N}\omega_i,
\label{eq:harmboltzmann}
\end{equation}
where $Z$ is the partition function, $k_B$ is Boltzmann's constant,
$T$ is the temperature, the elements of the matrix $\left|{\bf \Omega}\right|^2={\rm diag} (\omega_1^2,\ldots,\omega_{3N}^2)$ are the eigenvalues of
${\bf H}$, and the columns of the matrix ${\bf V}$ are the
eigenvectors of ${\bf H}$. To calculate $P({\bf x}_1)$ we define the
submatrices ${\bf H}_1$, ${\bf H}_2$, and ${\bf G}$ as follows:
\begin{eqnarray}
{\bf H\,x} = \left(
\begin{array}{cc}
{\bf H}_1 & {\bf G} \\
{\bf G}^\dag & {\bf H}_2
\end{array} 
\right)
\left(
\begin{array}{c}
{\bf x}_1 \\
{\bf x}_2
\end{array}
\right)
=
\left(
\begin{array}{ccc}
{\bf H}_1 {\bf x}_1 & + & {\bf G} {\bf x}_2 \\
{\bf G}^\dag {\bf x}_1 & + & {\bf H}_2 {\bf x}_2
\end{array} 
\right).
\label{eq:Hdecomp}
\end{eqnarray}
${\bf H}_1$ couples coordinates from ${\bf x}_1$; ${\bf H}_2$ couples coordinates from ${\bf x}_2$; and ${\bf G}$ couples coordinates between ${\bf x}_1$ and ${\bf x}_2$. Eq.~(\ref{eq:harmboltzmann}) now can be expressed as
\begin{equation}
P({\bf x})=Z^{-1}e^{{-{\bf x}^\dag {\bf H}\,{\bf x}}\over{2 k_B T}}=(2\pi k_B T)^{-3N/2}e^{{-\left|{\bf \bar{\Omega}} {\bf \bar{V}}^\dag {\bf x}_1\right|^2 - \left|{\bf \Lambda} {\bf U}^\dag {\bf x}_2 + {\bf \Lambda}^{-1}{\bf U}^\dag {\bf G}^\dag {\bf x}_1 \right|^2} \over {2 k_B T}}\prod_{i=1}^{3N}\omega_i,
\label{eq:harmsub}
\end{equation}
where the diagonal elements of the matrix $\left|{\bf \Lambda}\right|^2={\rm
diag}(\lambda_1^2,\ldots,\lambda_{3N_1}^2)$ and the columns of the
matrix ${\bf U}$ are the eigenvalues and eigenvectors of ${\bf H}_2$,
and the diagonal elements of the matrix $\left|{\bf \bar{\Omega}}\right|^2={\rm
diag}(\bar{\omega}_1^2,\ldots,\bar{\omega}_{3N_1}^2)$ and the columns of the
matrix ${\bf \bar{V}}$ are the eigenvalues and eigenvectors of a matrix ${\bf \bar{H}}$ defined as 
\begin{equation}
{\bf \bar{H}}={\bf H}_1 - {\bf G}{\bf H}_2^{-1}{\bf G}^\dag = {\bf \bar{V}}\left|{\bf \bar{\Omega}}\right|^2{\bf \bar{V}}^\dag.
\label{eq:Hbar}
\end{equation}
Eq.~(\ref{eq:Hbar}) is equivalent to an equation independently derived to
study local vibrations in the nucleotide-binding pockets of myosin and
kinesin \cite{Zheng05}. Performing the integral in
Eq.~(\ref{eq:px1_general}) leads to the desired equation for
$P({\bf x}_1)$:
\begin{equation}
P({\bf x}_1) = (2\pi k_B T)^{-3N_1/2}e^{{-\left|{\bf \bar{\Omega}}{\bf \bar{V}}^\dag {\bf x}_1 \right|^2} \over {2 k_B T}} \prod_{i=1}^{3N_1}\bar{\omega}_i.
\label{eq:margin}
\end{equation}

Now consider the problem of optimal selection of the parameters
$\Gamma$ of a coarse-grained model of protein dynamics. Let ${\bf
x}_\alpha$ be the coordinates of the $N_\alpha$ alpha-carbons in an an
all-atom model, and ${\bf x}_\alpha^{(\Gamma)}$ be the same
coordinates in the coarse-grained model. We define the optimal
coarse-grained model as the one for which the Kullback-Leibler
divergence between $P^{(\Gamma)}({\bf x}_\alpha)$ and $P({\bf
x}_\alpha)$ is minimal, {\em i.e.}, for which $\Gamma$ is chosen such
that
\begin{equation}
D_{{\bf x}_\alpha}^{(\Gamma)}=\int d^{3N_\alpha}{\bf x}_\alpha \,
P^{(\Gamma)}({\bf x}_\alpha)\ln {P^{(\Gamma)}({\bf x}_\alpha) \over
P({\bf x}_\alpha)}
\label{eq:dkl}
\end{equation}
is minimal. We previously calculated an analytic expression for
equations like Eq.~(\ref{eq:dkl}) when $P({\bf x}_\alpha)$ and
$P^{(\Gamma)}({\bf x}_\alpha)$ are both governed by harmonic
vibrations \cite{Ming05}:

\begin{equation}
D_{{\bf x}_\alpha}^{(\Gamma)}=\sum_{i=1}^{3N_\alpha}\left(\ln {\omega_{i}^{(\Gamma)} \over \bar{\omega}_{i}} + {1 \over {2 k_B T}}\bar{\omega}_{i}^2 \left| {\bf \bar{v}}^\dag_i \Delta{\bf x}_{\alpha,{0}}\right|^2 + {1 \over 2}\sum_{j=1}^{3N_\alpha}{\bar{\omega}_{j}^2 \over {\omega^{(\Gamma)}_{i}}^2}\left|{{\bf v}^{(\Gamma)}_i}^\dag {\bf \bar{v}}_j\right|^2- {1 \over 2}\right).
\label{eq:cgd}
\end{equation}
In Eq.~(\ref{eq:cgd}), ${\omega^{(\Gamma)}_i}^2$ and ${{\bf
v}^{(\Gamma)}_i}$ are the eigenvalue and eigenvector of mode $i$ of
the coarse-grained model; ${\bar{\omega}_i}^2$ and ${{\bf \bar{v}}_i}$
are the $i^{\rm th}$ eigenvalue and eigenvector of the matrix ${\bf
\bar{H}}$ calculated for the alpha-carbon atoms of the all-atom model
(Eq.~(\ref{eq:Hbar})), and $\Delta{\bf x}_{\alpha,{0}}={\bf
x}^{(\Gamma)}_{\alpha,0}-{\bf x}_{\alpha,0}$ is the difference between
the equilibrium coordinates of the coarse-grained and all-atom
models. An optimal coarse-grained model of harmonic vibrations is thus
one with parameters $\Gamma$ such that $D^{(\Gamma)}_{{\bf x}_\alpha}$
calculated using Eq.~(\ref{eq:cgd}) is minimal.

In the ENM \cite{Atilgan01}, interacting alpha-carbon atoms are
connected by springs aligned with the direction of atomic
separation. Following the Tirion model of harmonic vibrations
\cite{Tirion96}, each spring has the same force constant $\gamma$. For
a given interaction network, the eigenvectors ${\bf v}^{(\gamma)}_i$
are independent of $\gamma$, and each eigenvalue
${\omega^{(\gamma)}_i}^2$ is proportional to $\gamma$. The value of
$\gamma$ at which $D^{(\gamma)}_{{\bf x}_\alpha}$ is minimal may be
calculated using Eq.~(\ref{eq:cgd}):
\begin{equation}
\gamma={1 \over
{3N_\alpha}}\sum_{i=1}^{3N_\alpha}\sum_{j=1}^{3N_\alpha}
{\bar{\omega}_j^2 \over a_i^2}\left|{\bf v}_i^{(\gamma) \dag}{\bf
\bar{v}}_j\right|^2.
\label{eq:dmin}
\end{equation}
The proportionality constants $a_i^2={\omega^{(\gamma)}_i}^2/\gamma$
are determined from the eigenvalue spectrum calculated using an
arbitrary value of $\gamma$ (because the eigenvalues
$\omega^{(\gamma)2}_i$ are proportional to $\gamma$, the constants
$a_i^2$ are independent of $\gamma$). It is easily shown that the
third and fourth terms of Eq.~(\ref{eq:cgd}) cancel when $\gamma$
assumes the value given by Eq.~(\ref{eq:dmin}).

The interaction network in an elastic network
model is generated by enabling interactions only between pairs of
atoms separated by a distance less than or equal to a cutoff distance
$r_c$. To optimize the model,
the value of $r_c$ for which
$D^{(\gamma)}_{{\bf x}_\alpha}$ is minimal is numerically estimated, 
using values of $\gamma$
from Eq.~(\ref{eq:dmin}). 

As a test case for optimization, we developed a coarse-grained model
of bovine trypsinogen from an all-atom model (223 amino acids obtained
from PDB entry 4TPI \cite{Bode84}).  CHARMM was used for all-atom
simulations using the CHARMM22 force field with default parameter
values. HBUILD was used to generate hydrogen positions, and the energy
was initially minimized using 2000 steps of relaxation by the
adopted basis Newton-Raphson method, gradually reducing the weight of
a harmonic restraint to the crystal-structure coordinates. The final
minimized structure was obtained through vacuum minimization until a
gradient of $10^{-7}$ Kcal/mol\,\AA~was achieved, and the Hessian {\bf
H} was calculated in CHARMM. The coordinates of the elastic network
model were taken from the alpha-carbon coordinates of the minimized
all-atom model.

\begin{figure}
\includegraphics[width=3.0in]{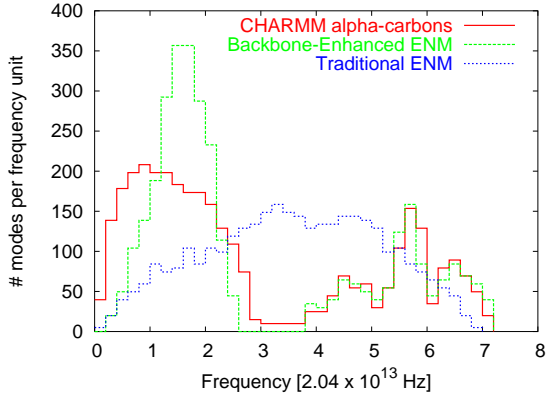}
\caption{Density-of-states distribution for all-atom and elastic
network models of trypsinogen. Frequency units are $({\rm Kcal}/ {\rm
mol} \, {\rm \AA}^2 \, m_p)^{1/2} = 2.04 \times 10^{13} \, {\rm Hz}$,
where $m_p$ is the proton mass. Densities were estimated by counting
the number of modes in bins of width 0.2, and normalizing the integral
to 663, which is the total number of non-zero modes. The ENM ({\em
dotted blue}) does not reproduce the bimodal distribution from the
all-atom model ({\em solid red}); however, the BENM recovers the
bimodal distribution ({\em dashed green}).}
\label{fig:freqs}
\end{figure}

The alpha-carbon vibrations of the all-atom model were calculated by
diagonalizing ${\bf \bar{H}}$ from Eq.~(\ref{eq:Hbar}). Interestingly,
the distribution of the density-of-states for the vibrations is
bimodal (Fig.~\ref{fig:freqs}) with 2/3 of the frequencies in the
low-frequency spectrum and 1/3 of the frequencies in the
high-frequency spectrum. Calculation of the density-of-states
distribution from other globular proteins yields bimodal patterns with
a similar 2:1 ratio between the numbers of low- and high-frequency
modes (unpublished results).

The best elastic network model of trypsinogen was obtained using a
cutoff distance $r_c$ of approximately 7.75~\AA, for which the optimal
value of $\gamma$ is 53.4~Kcal/mol\,\AA$^2$, yielding a value of
$D_{{\bf x}_\alpha}=312.9$ in a sharp minimum with respect to
$r_c$. The density-of-states distribution for the elastic network
model is unimodal, unlike that for the all-atom model
(Fig.~\ref{fig:freqs}).

Although the ENM treats all alpha-carbon pairs equally, 
the distribution of distances
between successive alpha-carbons along the protein
backbone is known to be tightly
centered about 3.8~\AA. In addition, two of the six alpha-carbons
nearest to a typical alpha-carbon are backbone neighbors, which might
explain why 1/3 of the CHARMM-derived modes have significantly higher
frequencies than the others. We therefore wondered whether the ENM
might be improved by enhancing interactions between backbone
neighbors.

Indeed, a more accurate coarse-grained model is obtained by using a
force constant enhanced by a factor of $\epsilon$ for interactions
between alpha-carbons that are neighbors on the backbone. Minimization
of $D_{{\bf x}_\alpha}$ for such a backbone-enhanced elastic network
model (BENM) with respect to $\epsilon$ and $r_c$ subject to
Eq.~(\ref{eq:dmin}) yields a model with $\epsilon=42$, $r_c=10.5$~\AA,
and $\gamma=4.26$~Kcal/mol\,\AA$^2$, resulting in a much lower value
$D_{{\bf x}_\alpha}=102.3$. The density-of-states distribution for
this model agrees quite well with that of the all-atom model
(Fig.~\ref{fig:freqs}), especially considering that the model is
optimized with respect to $D_{{\bf x}_\alpha}$, which does not
directly involve the density-of-states distribution. The agreement is
especially good for the high-frequency modes, suggesting that a
uniform force constant is a reasonable approximation for interactions
between alpha-carbons that are backbone neighbors. Furthermore, the
overlap $\sum_{i=1}^N\sum_{j=1}^N|{\bf v}_i^{(\gamma)\dag}\bar{\bf
v}_j|^2/N$ for the 223 highest-frequency modes is 0.99, indicating
that the spaces of the high-frequency eigenvectors are nearly
identical between the BENM and all-atom models. In contrast, the
low-frequency distribution of BENM states is narrower than that of the
all-atom model, indicating that a uniform force constant is a poorer
approximation for interactions between alpha-carbons that are not
backbone neighbors.

\begin{figure}
\includegraphics[width=3.0in]{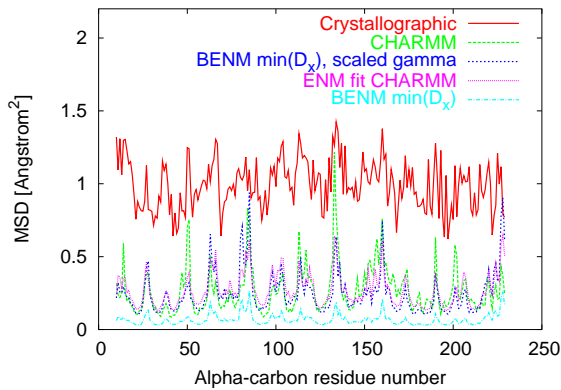}
\caption{Mean-squared displacements of alpha-carbon positions for
trypsinogen residues 10--229 obtained from normal-modes simulations
using CHARMM ({\em dashed green}), a BENM with parameters that
minimize $D_{{\bf x}_\alpha}$ with respect to CHARMM ({\em dotted
blue}), the same BENM but with $\gamma$ adjusted to better agree with
CHARMM MSDs ({\em fine-dotted magenta}), and an ENM with parameters
adjusted to agree with CHARMM MSDs ({\em dash-dotted cyan}). Values
were calculated at $T=300$~K using the Equipartition Theorem. Harmonic
vibrations at thermal equilibrium are known to inadequately
model crystallographic MSDs, which include other
sources of disorder ({\em solid red})
\cite{Go83}.}
\label{fig:flucts}
\end{figure}

Both the BENM and the ENM yield patterns of alpha-carbon MSDs that are
similar to that of the all-atom model (Fig.~\ref{fig:flucts}). Because
there are fewer low-frequency BENM modes than low-frequency CHARMM
modes (Fig.~\ref{fig:freqs}), the BENM MSDs are consistently smaller
than the CHARMM MSDs; however, the BENM MSDs may be improved by
selecting $\gamma=1.2$~Kcal/mol\,\AA$^2$
(Fig.~\ref{fig:flucts}). These improved MSDs come at the cost of a
higher value of $D_{{\bf x}_\alpha}=528.4$, and a change in the
frequency scale by a factor $(1.2/4.3)^{1/2}=0.53$, resulting in a
poor model of the density-of-states distribution. The ENM with
parameters that minimize $D_{{\bf x}_\alpha}$ exhibits poor MSDs (not
shown); however, an ENM with $r_c=15.4$~\AA\ and
$\gamma=0.4$~Kcal/mol\,\AA$^2$ yields MSDs that agree well with those
of the CHARMM model (Fig.~\ref{fig:flucts}). In agreement with
previous results using the ENM \cite{Atilgan01}, we confirmed that the
parameters of both the ENM and BENM may be adjusted to yield a
reasonable model of crystallographic MSDs (not shown).

Next consider the problem of quantifying allosteric effects in
proteins \cite{Ming05}. In allosteric regulation, molecular
interactions cause changes in protein activity through changes in
protein conformation. Although the importance of considering
continuous conformational distributions in understanding allosteric
effects was recognized by Weber \cite{Weber72}, theories of allosteric
regulation that consider continuous conformational distributions have
been lacking. We began to develop such a theory by defining the
allosteric potential as the Kullback-Leibler divergence $\bar{D}_{\bf
x}$ between protein conformational distributions before and after
ligand binding, and by calculating changes in the conformational
distribution of the full protein-ligand complex in the harmonic
approximation \cite{Ming05}. Here we use the expression for the
marginal distribution in Eq.~(\ref{eq:margin}) to calculate an equation
for the allosteric potential in the harmonic approximation, and apply
it to analyze allosteric mechanisms in trypsinogen.

Let
${\bf x}_p$ be the protein coordinates selected from the coordinates
${\bf x}$ of a protein-ligand complex. $P^\prime({\bf x}_p)$ and
$P({\bf x}_p)$ are the protein conformational distributions with and
without a ligand interaction. Eq.~(\ref{eq:margin}) enables
$P^\prime({\bf x}_p)$ to be calculated from the full conformational
distribution $P^\prime({\bf x})$ of the protein-ligand complex. The
equation for the allosteric potential in the harmonic
approximation follows from the theory developed in ref.~\cite{Ming05}:
\begin{equation}
\bar{D}_{{\bf x}}=\sum_{i=1}^{3N_p}\left(\ln
{\bar{\omega}^{\prime}_{i} \over {\omega}_{i}} + {1 \over {2 k_B
T}}{\omega}_{i}^2 \left| {\bf {v}}^\dag_i \Delta{\bf x}_{p,{0}}\right|^2 +
{1 \over 2}\sum_{j=1}^{3N_p}{{\omega}_{j}^2 \over
{\bar{\omega}^{\prime 2}_{i}}}\left|{{\bf \bar{v}}^{\prime \dag}_i}
{\bf {v}}_j\right|^2- {1 \over 2}\right).
\label{eq:ap}
\end{equation}
In Eq.~(\ref{eq:ap}), $\bar{\omega}^{\prime 2}$ and ${\bf
\bar{v}}^{\prime}_i$ are the $i^{\rm th}$ eigenvalue and eigenvector
of the matrix ${\bf \bar{H}}$ calculated for the protein atoms of the
protein-ligand complex, $\omega_i^2$ and ${\bf v}_i$ are the
eigenvalue and eigenvector of mode $i$ of the apo-protein, and $\Delta
{\bf x}_{p,0}={\bf x}^\prime_{p,0}-{\bf x}_{p,0}$ is the difference
between the equilibrium coordinates of the protein with and without
the ligand interaction. The term
$\sum_{i=1}^{3N_p}\ln{\bar{\omega}^\prime_i / \omega_i}$ is
proportional to the change in configurational entropy of the protein
releasing the ligand, and the term
$\sum_{i=1}^{3N_p}{\omega}_{i}^2 \left| {\bf {v}}^\dag_i \Delta{\bf
x}_{p,{0}}\right|^2 / 2 k_B T$ is proportional to the potential energy
required to deform the apo-protein into its equilibrium
conformation in the protein-ligand complex.

We used Eq.~(\ref{eq:ap}) to calculate changes in the configurational
distribution of local regions of trypsinogen upon binding bovine
pancreatic trypsinogen inhibitor (BPTI). BPTI binds in the active site
and exerts an allosteric effect, enhancing the affinity of trypsinogen
for Val-Val \cite{Bode79}. Alpha-carbon coordinates for 223 residues
were obtained from a crystal structure of trypsinogen in complex with
BPTI (residues 7--229 from PDB entry 4TPI \cite{Bode84}, including
theoretically modeled residues 7--9), and were used directly to
construct backbone-enhanced elastic network models of apo-trypsinogen
and the trypsinogen-BPTI complex. As suggested by the refined
trypsinogen model above, both models used $r_c=10.5$~\AA,
$\gamma=4.26$~Kcal/mol\,\AA$^2$, and $\epsilon=42$.

Local changes in the conformational distribution of trypsinogen were
analyzed by considering changes in the neighborhood of each
alpha-carbon atom. A neighborhood was defined by selecting the atom of
interest plus its five nearest neighbors, and the matrix ${\bf
\bar{H}}$ was calculated for these six atoms in the models both with
(yielding ${\bf \bar{H}}^\prime$) and without (yielding ${\bf
\bar{H}}$) the BPTI interaction. A local value of $\bar{D}_{\bf x}$
was obtained using the eigenvalues and eigenvectors of ${\bf
\bar{H}}^\prime$ and ${\bf \bar{H}}$ in a suitably modified version of
Eq.~(\ref{eq:ap}).

\begin{figure}
\includegraphics[width=3in]{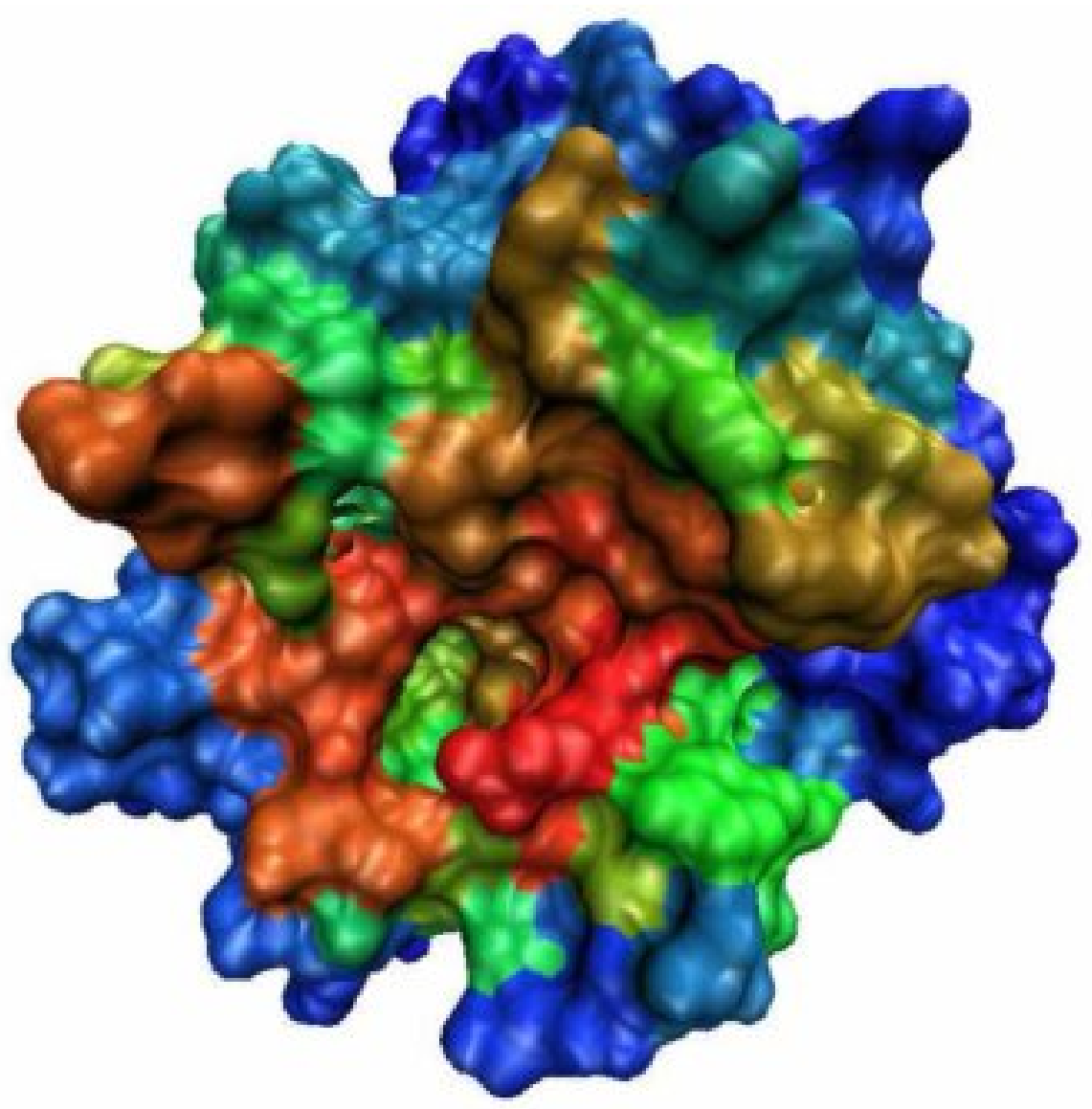}
\includegraphics[width=3in]{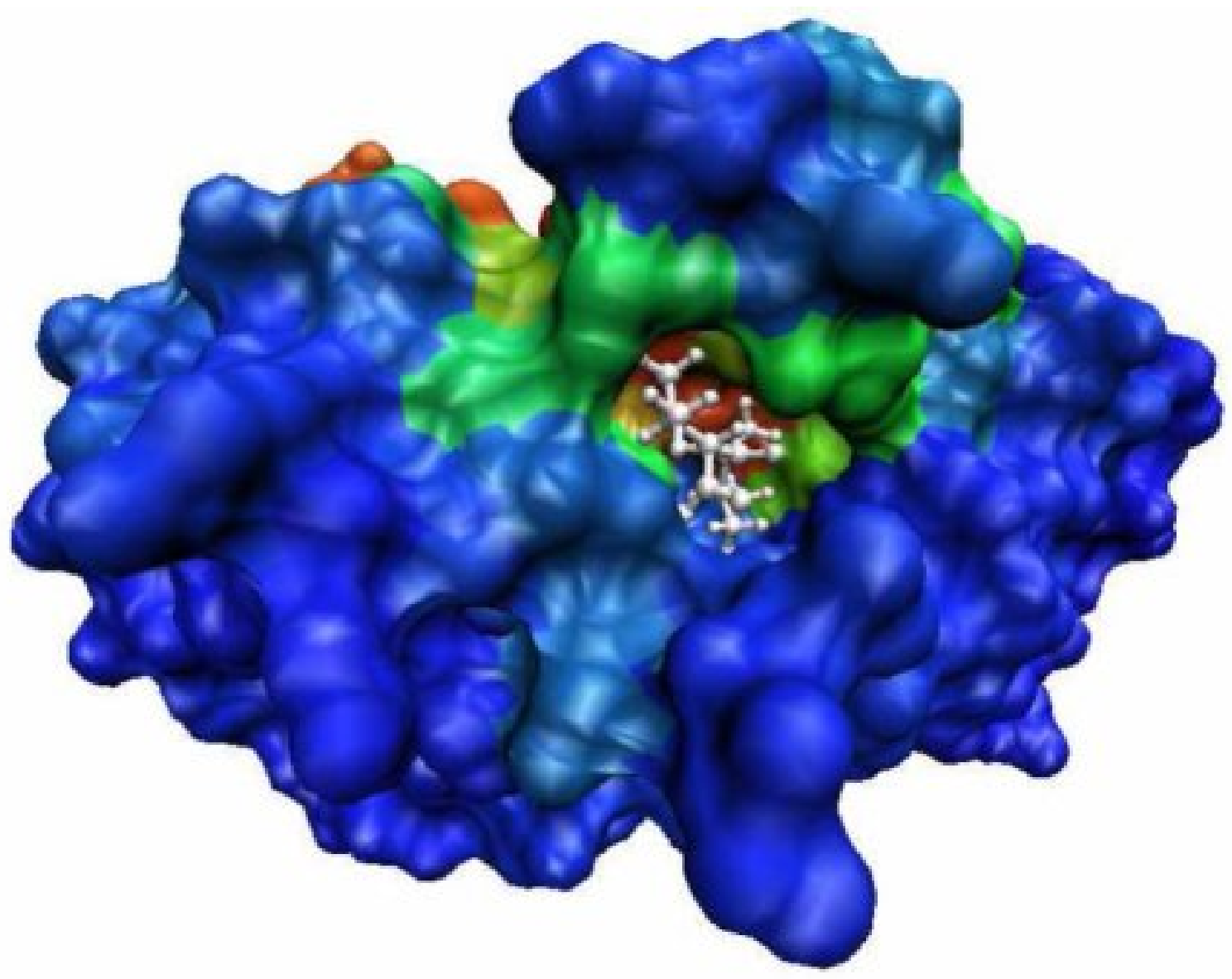} 
\caption{Visualization of local sites on the surface of trypsinogen
that exhibit a large change in the conformational distribution upon
binding BPTI. Values of $\bar{D}_{\bf x}$ are mapped to a logarithmic
temperature scale, with red coloring indicating large values. Changes
are large both in the BPTI-binding site ({\em left}) and in the
Val-Val binding site ({\em right}). There is a $90^\circ$ rotation
about the x-axis between the left and right panels.}
\label{fig:trypsinogen}
\end{figure}

Not surprisingly, we found that the local values of $\bar{D}_{\bf x}$
were relatively large in the neighborhood of the BPTI-binding site
(Fig.~\ref{fig:trypsinogen}, {\em left panel}). Values of
$\bar{D}_{\bf x}$ elsewhere on the surface were smaller, with one
interesting exception: values in the Val-Val binding site were
comparable to those in the BPTI-binding site
(Fig.~\ref{fig:trypsinogen}, {\em right panel}).

We also calculated local values of $\bar{D}_{\bf x}$ for the Val-Val
interaction, which causes the crystal structure of trypsinogen to
resemble that of active trypsin \cite{Bode78,Bode84}. We found that
values were relatively large in the neighborhood of Ser 195, which is
the key catalytic residue for trypsin and other serine proteases: the
value of $\bar{D}_{\bf x}$ in this neighborhood was 40$^{\rm th}$
highest of 223 residues in the crystal structure; 11$^{\rm th}$ of all
residues not directly interacting with the Val-Val in the model; the
highest of all residues located at least as far as Ser 195 is from the
Val-Val ligand; and greater than that for 20 of 60 residues located
closer to the ligand. Calculations for both the BPTI interaction and
the Val-Val interaction therefore indicate that there is a relatively
strong communication between the regulatory and active sites of
trypsinogen.

Considering models beyond the ENM and BENM (and even models beyond
proteins), the theory presented here leads to a general prescription
for modeling harmonic vibrations using coarse-grained models of
materials. To optimally model the all-atom conformational
distribution, always use an energy scale for interactions that
eliminates the discrepancy due to differences in the eigenvectors
(Eq.~(\ref{eq:dmin})), and select the 
coarse-grained model for which the entropy of the conformational
distribution is the largest (first term of Eq.~(\ref{eq:cgd})).

Although traditional elastic network models can explain
characteristics of the functions and dynamics of proteins
\cite{Yang05}, the present study shows that they provide a poor
approximation to the conformational distribution calculated from
all-atom models of harmonic vibrations of proteins. Model accuracy is
significantly improved by using a backbone-enhanced elastic network
model, which strengthens interactions between atoms that are nearby in
terms of covalent linkage. Although the backbone-enhanced model
appears to accurately capture the high-frequency alpha-carbon
vibrations of an all-atom model, the model less accurately captures
the slower, large-scale harmonic vibrations (which in turn are known
to poorly approximate the full spectrum of highly nonlinear,
large-scale protein motions).

We also find that the allosteric potential is a useful tool for
computational analysis of allosteric mechanisms in proteins. Using
calculations of the allosteric potential, communication between the
regulatory and active sites of trypsinogen was observed in a purely
mechanical, coarse-grained model of protein harmonic vibrations that
does not consider mean conformational changes or amino-acid
identities, supporting prior arguments for the possibility of
allostery without a mean conformational change \cite{Cooper84}. It will
be interesting to perform similar analyses on a wide range of all-atom
and coarse-grained models of protein vibrations, and to use more
realistic calculations of free-energy landscapes \cite{Garcia01} to
more accurately model changes in protein conformational distributions.

This work was supported by the US Department of Energy.

\bibliography{PRL}

\end{document}